\documentclass[journal]{IEEEtran}
\usepackage{mathtools,amssymb,bm}
\usepackage{amsmath}
\usepackage{cite}
\usepackage{graphicx}
\usepackage{epstopdf}
\usepackage{cite}
\usepackage{color}
\usepackage{caption}
\usepackage[footnotesize]{subfigure}
\usepackage{hyperref}
\begin{document}
\title{Eliminating Impulsive Noise in Pilot-Aided OFDM Channels via Dual of Penalized Atomic Norm }
\author{Iman Valiulahi, Farzad Parvaresh, and  Ali Asghar Beheshti
	\thanks{I. Valiulahi and A. Beheshti are with the School of Electrical Engineering, Iran University of Science \& Technology (IUST), Tehran, Iran. F. Parvaresh is with the Department of Electrical Engineering, University of Isfahan, Isfahan, Iran (email: f.parvaresh@eng.ui.ac.ir) and also School of Mathematics, Institute for Research in Fundamental
		Sciences (IPM), P.O. Box: 1395–5746, Tehran, Iran. F. Parvaresh was supported in part by a grant from IPM (No. 96680423). }
}%
\maketitle

\begin{abstract}
	 In this paper, we propose a novel estimator for pilot-aided orthogonal frequency division multiplexing (OFDM) channels in an additive Gaussian and impulsive perturbation environment.  Due to sensor failure which might happen because of man-made noise, a number of measurements in high rate communication systems is often corrupted by impulsive noise. High power impulsive noise is generally an obstacle for OFDM systems as valuable information will be completely lost. To overcome this concern, an objective function based on a penalized atomic norm minimization (PANM) is provided in order to promote the sparsity of time dispersive channels and impulsive noise. The corresponding dual problem of the PANM is then converted to tractable semidefinite programming. It has shown that one can simultaneously estimate the time dispersive channels in a continuous dictionary and the location of impulsive noise using the dual problem. Several numerical experiments are carried out to evaluate the performance of the proposed estimator.      
\end{abstract}

\IEEEpeerreviewmaketitle

\vspace{-0.3cm}
\section{Introduction}
Over the past few years, there has been a growing interest in using orthogonal frequency division multiplexing (OFDM) for high rate communication systems because of its large number of closely spaced orthogonal sub-carriers, which lead to parallel transmission. One of the most important issues facing wireless communication systems is to transmit data over channels with man-made noise which is common in the urban environments. The man-made noise might be occurred by ignition systems, heavy power lines, and current switches. The man-made noise does not follow Gaussian noise structure and must be modeled by impulsive noise \cite{blackard1993measurements}.  Impulsive noise with high amplitude and wide bandwidth occurs rapidly. The long duration and high repetition rate of impulsive noise can effectively destroy a subset of measurements at the receiver side, leading to significantly reduction in the performance of communication systems \cite{blackard1993measurements}.

 Generally speaking, OFDM systems are more robust against impulsive noise than single carrier systems. The longer symbol duration in the OFDM systems helps to simultaneously spread impulsive noise effects on OFDM subcarriers \cite{zhidkov2008analysis}. Nevertheless, strong interferences with high power, which can be found in industrial environments, might diminish system performance and even break the radio communication link in the OFDM systems \cite{zhidkov2008analysis}.

In the OFDM systems,  due to the fact that the number of strong paths is much less than total paths, the channel is technically sparse \cite{simeone2004pilot}. Thus, one can use super-resolution methods such as MUSIC \cite{schmidt1986multiple} to estimate the channel parameters. However, subspace decomposition approaches are insufficient in the presence of impulsive noise as they are intended to handle Gaussian-type noise. In \cite{zhidkov2008analysis}, an algorithm for the OFDM systems proposed when the received signal is corrupted by the additive Gaussian and impulsive noise. In the proposed approach, a new blanking nonlinearity block is needed to use before the OFDM demodulator \cite{zhidkov2008analysis}.  

Regarding the sparsity of the channel, one can leverage the results of compressed sensing (CS) for estimating the OFDM channels \cite{cheng2013channel}. $\ell_{1}$ minimization (LM), however, suffers from basis mismatch because of its inherent discretization \cite{chi2011sensitivity}. Indeed, the time dispersive channels belong to a continuous dictionary, whereas, in the grid-based CS, they are assumed to lie on a fine grid \cite{pejoski2015estimation}. Hence, mismatching between the reconstructed and inherent time dispersive channels is undeniable \cite{chi2011sensitivity}.  In order to detect the sparse sources in a continuous domain using a partial observation, one can use atomic norm minimization (ANM) \cite{tang2013compressed}. ANM is then used for estimating the OFDM channel in \cite{pejoski2015estimation}. To see a significant comparison between ANM and conventional approaches such as MUSIC, one might take a look at \cite{bhaskar2013atomic}. Also, in \cite{fu2018quantized}, Cramer-Rao lower bound for quantized super resolution based on ANM is developed.
 
 Here, we suggest a novel, simple estimator for the pilot-aided OFDM channel using the penalized atomic norm minimization (PANM) when the measurements are corrupted by the additive Gaussian and impulsive noise. It has shown that one can estimate both the time dispersive channels in a continuous dictionary and the location of impulsive noise using the dual problem of the PANM. The results of positive trigonometric polynomial (PTP) theory  \cite{dumitrescu2017positive} is then used to convert the dual problem to linear matrix inequalities, which is solvable in polynomial time.  It is worth mentioning that the location of impulsive noise and the sparse time dispersive channels are obtained by the magnitude of the dual solution and the dual polynomial, respectively. When the time dispersive channels are determined, one can use the well-known least squares (LS) methods to recover the gain of each path. Notice that PANM only depends on the number of impulsive noise, time dispersive channels and pilots as shown in phase transition graphs in Section \ref{section.simulation}, whereas, a sophisticated analysis of impulsive noise power is needed for the conventional estimator such as \cite{zhidkov2008analysis}. Finally, several MATLAB simulations have been done to investigate the proposed estimator's performance. 
 
The paper is organized as follows: problem formulation is given in Section \ref{section.model}. In Section \ref{section.conicgeometry}, the primal problem of the PANM is proposed. In Section \ref{DualProblem}, the dual problem and a strategy for recovering the channel parameters and localizing impulsive noise are brought. In ُSection \ref{section.simulation}, numerical experiments are presented to support the main results. Finally, the paper is concluded in Section \ref{section.concultion}.

Throughout the paper, scalars are denoted by lowercase letters, vectors by lowercase boldface letters, and matrices by uppercase boldface letters. The $k$th element of the vector $\bm{x}$ is denoted by $x(k)$. The absolute value of a scalar, the element-wise absolute value of a vector and the cardinality of a set are shown by $|\cdot|$. The infinity norm is $\|\bm{z}\|_{\infty}=\underset{k}{\max}~|z(k)|$. In addition, the $\|\cdot\|_{1}$ and $\|\cdot\|_{2}$ are reserved for $\ell_{1}$ and $\ell_{2}$ norms, respectively. The operator $\langle\cdot,\cdot \rangle_{\mathbb{R}}$  stands for the real part of the inner product of two vectors. The operators $\text{Tr}(\cdot)$, $(\cdot)^{H}$ and $\ast$ are used to denote the trace of a matrix, the Hermitian of a vector and the convolution operator, respectively. To show that $\bm{A}$ is a semidefinite matrix we write $\bm{A}\succeq 0$.

\section{Problem Formulation}\label{section.model}
The aim of this section is to formulate a pilot-aided OFDM system with $N$ sub-carriers and a cycle prefix length of $L_{cp}$ when a subset of the received signal is corrupted by the additive impulsive and Gaussian noise. After coding and modulation, the bits of information are passed through a serial to parallel converter to obtain frequency domain symbols. These symbols are subjected to inverse discrete Fourier transform (IDFT) to produce the time domain symbols. A cycle prefix length of $L_{cp}$ is added. After parallel to serial conversion, the filter $g_T(\tau)$ is used to shape the OFDM symbols in order to achieve the desired signal to transmit through a multi-path environment with $s$ scatters. The base-band model of the channel can be written as a sparse combination of $s$ Dirac functions as below \cite{taubock2010compressive}:
              \begin{align}\label{1}
h(\tau)=\sum_{k=1}^{s}\alpha_{k}\delta(\tau-\tau_{k}),
\end{align}
 where $\alpha_{k}$ and $\tau_{k}$ are the complex amplitude and the real delay spread for the $k{th}$ path, respectively. The function $\delta(\tau-\tau_{k})$ denotes the Dirac delta function located at $\tau_{k}$. Under the assumption of the block-fading, the channel parameters are assumed to be constant over a block of data. At the receiver side, the contaminated signal by the additive Gaussian and impulsive noise is subjected to the matched filter $g_{R}(\tau)$. Assume that the accurate timing synchronization occurs, after serial to parallel conversion, when the cycle prefix is removed, the N-point DFT of the received signal are passed through a parallel to serial conversion. Thus, we have
  \begin{align}\label{model}
y_{n}(n)=h_{0}(n)x(n)+w_{1}(n)+z_{1}(n), \: n \in \mathcal{N}=\{1,\cdots, N\},\hspace{-0.1cm}
  \end{align}
    where $x(n)$ is the associated data (pilot) of the ${nth}$ sub-carrier, $\bm{w}_{1} \in \mathbb{C}^{ N \times 1}$ is the Gaussian noise vector whose elements modeled as i.i.d. zero mean circularly symmetric complex Gaussian variable with the variance $\sigma_{n}^{2}$, $\bm{z}_{1} \in \mathbb{C}^{ N \times 1}$ is used to denote the impulsive noise vector. Due to the fact that filters $g_{T}(\tau)$ and $g_{R}(\tau)$ are band-limited \cite{morelli2001comparison}, one can sample $h_{0}(n)$ from $
h_{0}(n)=g(n)h(n), ~~ n \in \mathcal{N}$, where $g(n)$ can be sampled from DTFT of the discretized $g(\tau)=g_{T}(\tau)\ast g_{R}(\tau)$ and 
  \begin{align}\label{Channel}
 h(n)=\sum_{k=1}^{s}\alpha_{k}e^{-j2\pi\tfrac{n}{N}\tfrac{\tau_{k}}{T_{s}}},
  \end{align}
where $T_{s}$ is the sampling interval. In practice, $T_{g}$, which is the duration of the significant part of $g(\tau)$, is set  $T_{g}=4 \div 6T_{s}$, so $ 0 \le \tau_{k} < L_{cp}T_{s}-T_{g}$ in order to satisfy the conditions about
the temporal support which \cite{simeone2004pilot} used to derive (\ref{model}).  Let consider the well-known pilot allocation scheme presented in \cite{simeone2004pilot}, one can define the integer $P=\tfrac{N}{D}$ for the integer $D$ such that $P=L_{cp}-\tfrac{T_{g}}{T_{s}}$. Due to the fact that $\tfrac{\tau_{k}}{NT_{s}} \in [0,\tfrac{P}{N})$, one can use $P$ pilots located at the sub-carrier position of $n_{p} \in \{0, \tfrac{N}{P}, \cdots, \tfrac{(P-1)N}{P}\}$ instead of $N$ sub-carriers \cite{simeone2004pilot}. Let use $n^{\prime}_{p}=\tfrac{P}{N}n_{p}\in \mathcal{J}=\{0,\cdots, P-1\}$ instead of $n_{p}$. Therefore, one can recast the observation model as below:  
\begin{align}\label{corruptedmeasurments}
 y(n^{\prime}_{p}\tfrac{N}{P})&= h(n^{\prime}_{p}\tfrac{N}{P})+ \tfrac{w_{1}(n^{\prime}_{p}\tfrac{N}{P})}{x(n^{\prime}_{p}\tfrac{N}{P})}+\tfrac{z_{1}(n^{\prime}_{p}\tfrac{N}{P})}{x(n^{\prime}_{p}\tfrac{N}{P})}\nonumber\\&=\sum_{k=1}^{s}\alpha_{k} e^{-j2\pi\tfrac{\tau_{k}}{PT_{s}}n^{\prime}_{P}}
 +w(n^{\prime}_{p}\tfrac{N}{P})+z(n^{\prime}_{p}\tfrac{N}{P}), 
  \end{align}
  where $\bm{w}$ and $\bm{z}$ denote the Gaussian noise vector with variance $\sigma^{2}_{np}$ and impulsive noise vector with $r$ nonzero entires supported at $\Omega$, respectively. 
   \begin{figure*}[t]
  	\centering
  	\mbox{
  		\hspace{-1cm}\subfigure[]{\includegraphics[height=3.2cm,width=10cm]{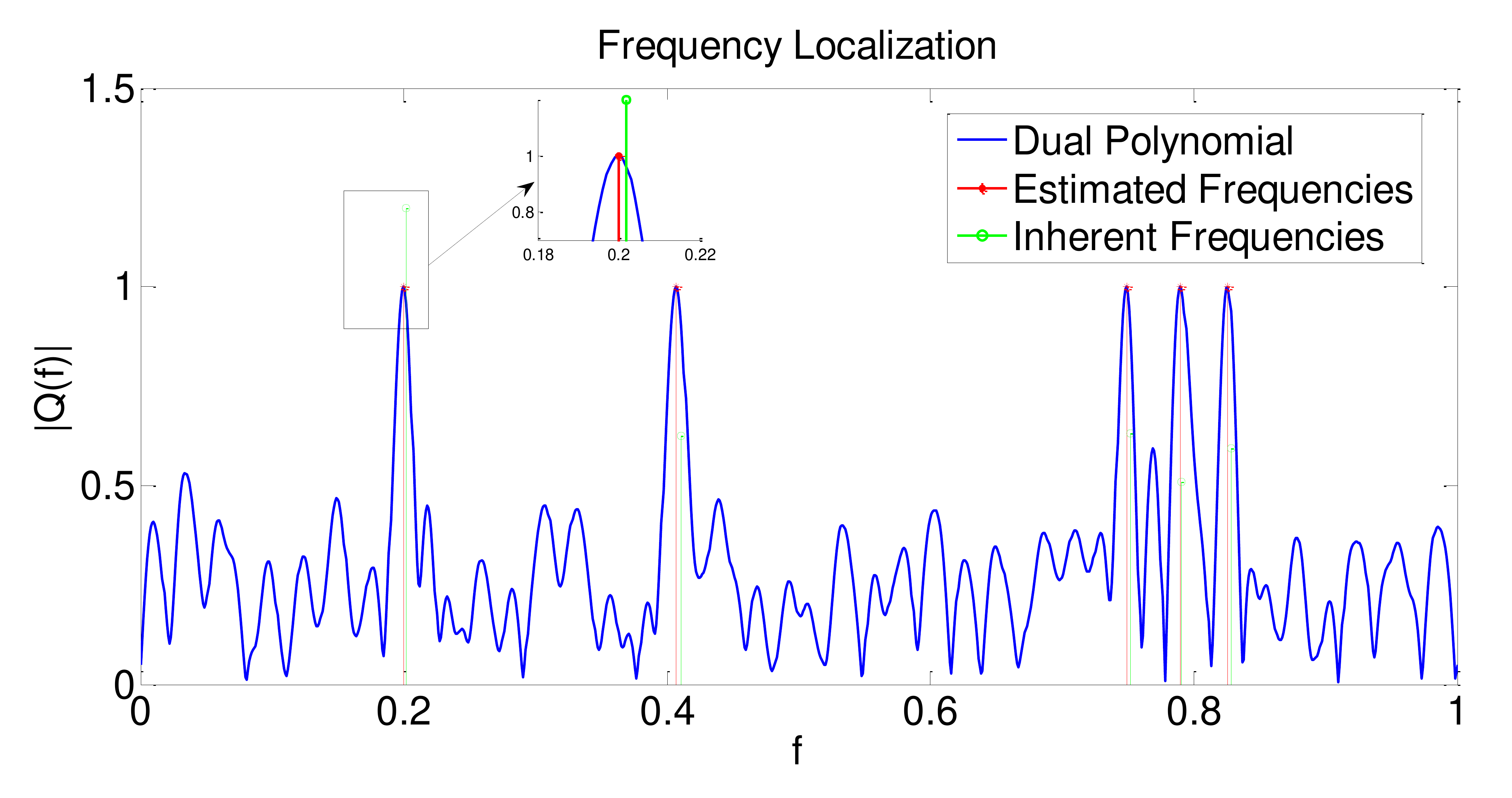}\label{fig.fre}}\hspace{-0.3cm}
  		\subfigure[]{\includegraphics[height=3.2cm, width=10 cm]{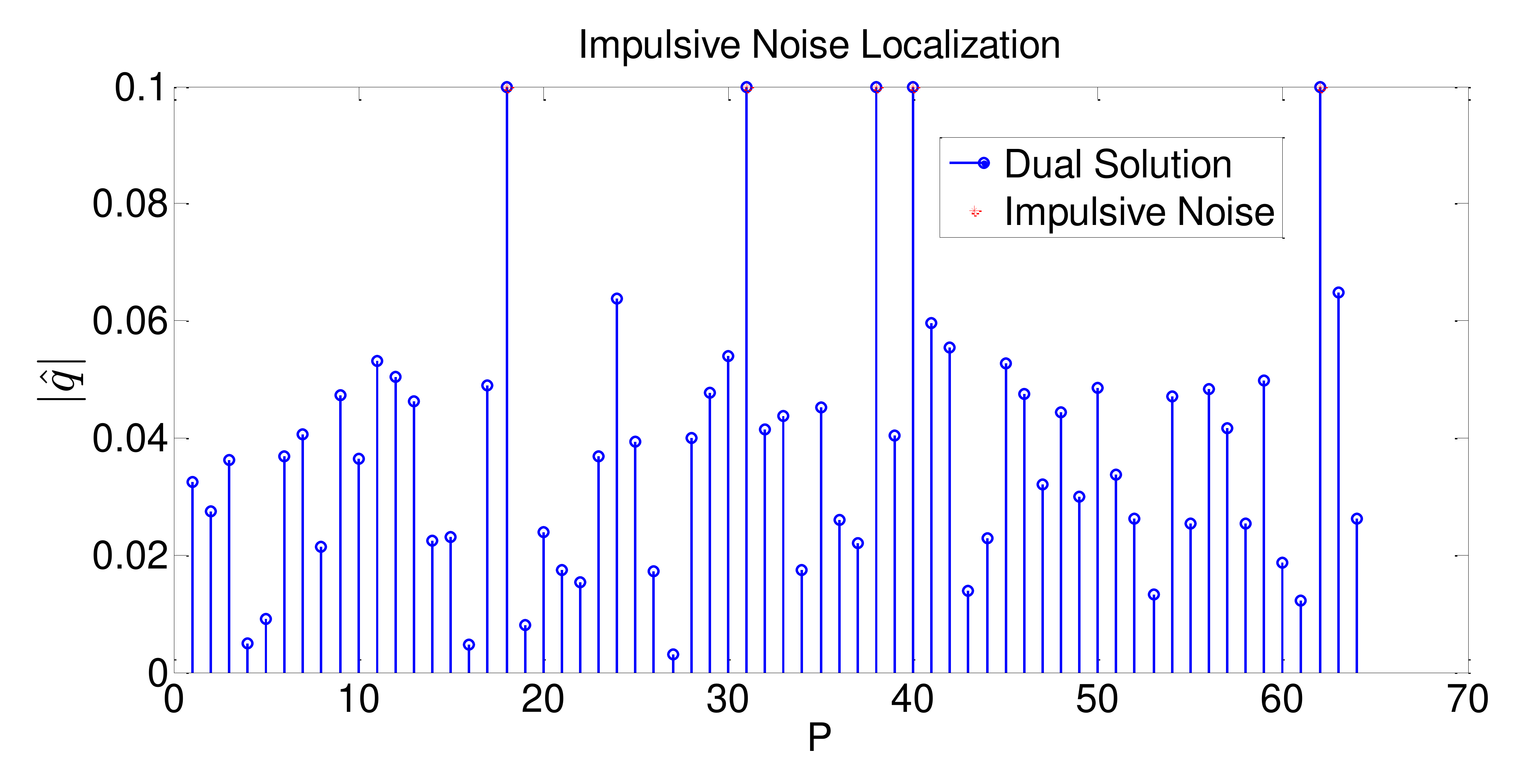}\label{fig.nois}}}\vspace{-0.3cm}
  	\caption{ Frequencies and impulsive noise localization using the dual problem (\ref{sdp}).\vspace{-0.5cm}}\label{fig.support}
  \end{figure*}
  The problem is to estimate $\tau_{k}$ and $\alpha_{k}$ for $k \in \{1,\cdots,s\}$ in (\ref{corruptedmeasurments}) using $P$ pilots when the observed vector $\bm{y}$ is contaminated by the additive Gaussian and impulsive noise.
  
\section{The Penalized Atomic Norm minimization}\label{section.conicgeometry}
 The primal problem of the PANM is provided in this section. Let first define $f_{k}:= \tfrac{1}{P}\tfrac{\tau_{k}}{T_{s}},~\forall k=\{1,\cdots,s\}$, thus $f_{k} \in [0,1]$.  Regarding this definition, we have $\sum_{k=1}^{s}\alpha_{k} e^{-j2\pi f_{k}n^{\prime}_{P}}=\int_{f \in [0,1]}e^{-j2\pi f n^{\prime}_{P}} \mu (f) df$ for $n^{\prime}_{p} \in \mathcal{J}$ where the integral on the right-hand side can be written in the matrix form as $\bm{h}=\mathcal{F}\mu$, where $\mathcal{F}$ is a linear operator that maps the measure $\mu (f)=\sum_{k=1}^{s}\alpha_{k}\delta(f-f_{k})$ to its lowest Fourier coefficients. Consequently, the problem is circumvented to estimate $f_{k}$ and $\alpha_{k}$ for $k \in \{1,\cdots,s\}$. One can find a sparse representation for the new $\bm{h}$ in the sinusoidal atoms $\bm{a}(f,\phi) \in \mathbb{C}^{P\times 1}$, $a(f,\phi)(l)=e^{j\phi}e^{-j2\pi l f},~~l\in \mathcal{J}$  with parameters $f \in [0,1]$ and $\phi \in [0,2\pi)$ as $
\bm{h}=\sum_{k=1}^{s}|\alpha_{k}|\bm{a}(f_{k},\phi_{k}), ~~ \alpha_{k}:=|\alpha_{k}|e^{j \phi_{k}}
$. 
 The atomic norm on the set $\mathcal{A}$ is defined by 
$\|\bm{u}\|_{\mathcal{A}}:=\inf \{t> 0: \bm{u}\in t\mathrm{conv}(\mathcal{A}) \}, 
$ where $\mathcal{A}:=\{\bm{a}(f,\phi): f\in[0,1), \phi \in [0,2\pi)\}$, the set $\mathrm{conv}(\mathcal{A})$ denotes the unite ball of the convex hull of $\mathcal{A}$, and $t$ is a constant variable.

 Let recast the corrupted measurement model (\ref{corruptedmeasurments}) in the matrix form as below:
\begin{align}\label{measutmentmatrixform}
\bm{y}=\bm{h}+\bm{w}+\bm{z},
\end{align} 
where $\bm{h}$, $\bm{w}$, and $\bm{z} \in \mathbb{C}^{P\times 1} $. Due to the fact that the observed signal is the sum of two sparse vectors in the different domains, $\bm{h}$ and $\bm{z}$, which are corrupted by dense noise, $\bm{w}$, one can use a convex optimization problem to promote side tasks. Here, we are interested in an optimization problem which can achieve a low-dimensional presentation for the channel as well as consider the sparsity of impulsive noise and the power of Gaussian perturbation as its additional tasks. To achieve this goal, we propose the following convex optimization problem
\begin{align}\label{primalproblem}
\min_{\tilde{\bm{h}},\tilde{\bm{z}}}~&\|\hat{\bm{h}}\|_{\mathcal{A}}+\lambda \|\tilde{\bm{z}}\|_{1}\nonumber\\&
\text{subject~to}~\|\bm{y}-\tilde{\bm{h}}-\tilde{\bm{z}}\|_{2}\le \sigma^{2}_{np},
\end{align}
where $\lambda >0$ is a regularization parameter which makes a balance between the sparsity of $\bm{h}$ and $\bm{z}$. There exist two possible ways to implement this convex optimization problem. First, it is possible to leverage the proposed semidefinite characterization by Tang et al. in \cite{tang2013compressed} for $\|\cdot\|_{\mathcal{A}}$ and the observed set $\mathcal{J}$ as $
\|\bm{x}\|_{\mathcal{A}}=\min_{\bm{u},g}\bigg\{\tfrac{1}{2|\mathcal{J}|}\text{Tr}(\mathcal{T}(\bm{u}))+\tfrac{g}{2}\bigg|
\begin{bmatrix}
\mathcal{T}(\bm{u})&\bm{x}\\
\bm{x}^{H}& g 
\end{bmatrix}
\succeq 0
\bigg\},
$
where $\mathcal{T}(\bm{u})$ denotes the Toeplitz matrix with $\bm{u}$ as the first column and $g$ is a constant number. One can localize the locations of the spectral sources using the Vandermonde decomposition of Toeplitz matrices discussed in \cite{tang2013compressed}. 
Note that the pair $(\tilde{\bm{h}}, \tilde{\bm{z}})$ is the solution of (\ref{primalproblem}), which does not directly provide the estimation of the time dispersive channels. Thus, we prefer the second approach based on the corresponding dual of (\ref{primalproblem}) to simultaneously localize both the time dispersive channels and impulsive noise. 

\section{ The Dual of The Penalized Atomic Norm}\label{DualProblem}

 Without loss of generality, let first define $m:=\tfrac{P-1}{2}$ or $m:=\tfrac{P}{2}-1$ if $P$ odd or even, respectively, thus,  $\mathcal{J}=\{-m,\cdots,m\}$. One can define the dual of $\|\cdot\|_{\mathcal{A}}$ as below \cite{tang2013compressed}:
\begin{align}
\|\bm{q}\|^{*}_{\mathcal{A}}&:=\sup_{\|\bm{h}\|_{\mathcal{A}}\le 1}\langle \bm{q},\bm{h}\rangle_{\mathbb{R}}\nonumber\\
&=\sup_{f \in [0,1],\phi \in [0,2\pi)}\langle \bm{q},e^{j \phi} \bm{a}(f,0)\rangle_{\mathbb{R}}=\sup_{f\in [0,1]}|\langle\bm{q},\bm{a}(f,0)\rangle|,\nonumber
\end{align}
where the second equality comes from the fact that $\|\cdot\|_{\mathcal{A}}$ is convex and its maximum occurs in the boundary of its domain. Using Lagrangian theorem \cite{boyd2004convex}, one can obtain the associated dual problem of (\ref{primalproblem}) as below:
\begin{align}\label{dual}
\max_{\bm{q}} ~&\langle \bm{q},\bm{y}\rangle_{\mathbb{R}}-\sigma_{np}\|\bm{q}\|_{2}\nonumber\\
&\text{subject~to} ~~\|\bm{q}\|^{*}_{\mathcal{A}}\le 1,~~\|\bm{q}\|_{\infty}\le \lambda,
\end{align}
where $\bm{q}\in \mathbb{C}^{P\times 1}$ is the dual solution. Note that Slater conditions hold between the primal and dual problem \cite{boyd2004convex}, so there exists no duality gap \cite{boyd2004convex}. The infinite dimensionality of $\|\bm{q}\|^{*}_{\mathcal{A}}\le 1$ is generally an obstacle \cite{tang2013compressed}. The results of PTP theory \cite{dumitrescu2017positive}, however, can be used to tackle this issue.  This theory claims that the magnitude of the trigonometric polynomials can be controlled by linear matrix inequalities \cite{dumitrescu2017positive}. Using PTP theory, one can suggest a precise semidefinite programming formulation for (\ref{dual}) as below \cite{dumitrescu2017positive}:
\begin{align}\label{sdp}
\begin{split}
\underset{\bm{q},\bm{Q}_0}{\max} ~~&{\langle \bm{y},\bm{q}\rangle}_{\mathbb{R}}-\sigma_{np}\|\bm{q}\|_{2}\\\nonumber
&\mathrm {subject\,\,to }~ \mathrm{tr}[\bm{\Theta}_{k}\bm{Q}_0]=\delta_{k},\quad k\,\in\ \mathcal{J},
\end{split}\\
&\hspace{1.5cm}\begin{bmatrix}
\bm{Q}_0 & {\bm{q}} \\
\\\bm{q}^H & 1
\end{bmatrix}
\succeq 0,\hspace{0,5cm}\|\bm{q}\|_{\infty}\le \lambda,
\end{align}
where $\bm{Q}_0\in \mathbb{C}^{P\times P}$ is a Hermitian matrix such that $\bm{Q}_{0} \succeq 0$, $\bm{\Theta}_{k}\in\mathbb{C}^{P\times P} $ is an elementary Toeplitz matrix with ones on it's k-th diagonal and zeros else where. Notice that $k=0$ is associated with
the main diagonal, positive and negative values are reserved for the upper and lower diagonals, respectively.  $\delta_{0}=1$ and $\delta_{k}=0$ if $k\ne0$. After solving (\ref{sdp}), $\hat{\bm{q}}$ can be used to construct the dual polynomial as below: 
\begin{align}\label{dualpoly}
Q(f)=\sum_{k=-m}^{m}\hat{q}(k)e^{j2\pi f k}.
\end{align}
\begin{figure*}[t]
	\centering
	\vspace{-0.4cm}
	\mbox{	\hspace{-1cm}
		\subfigure[]{\includegraphics[height=3cm, width=9.5cm]{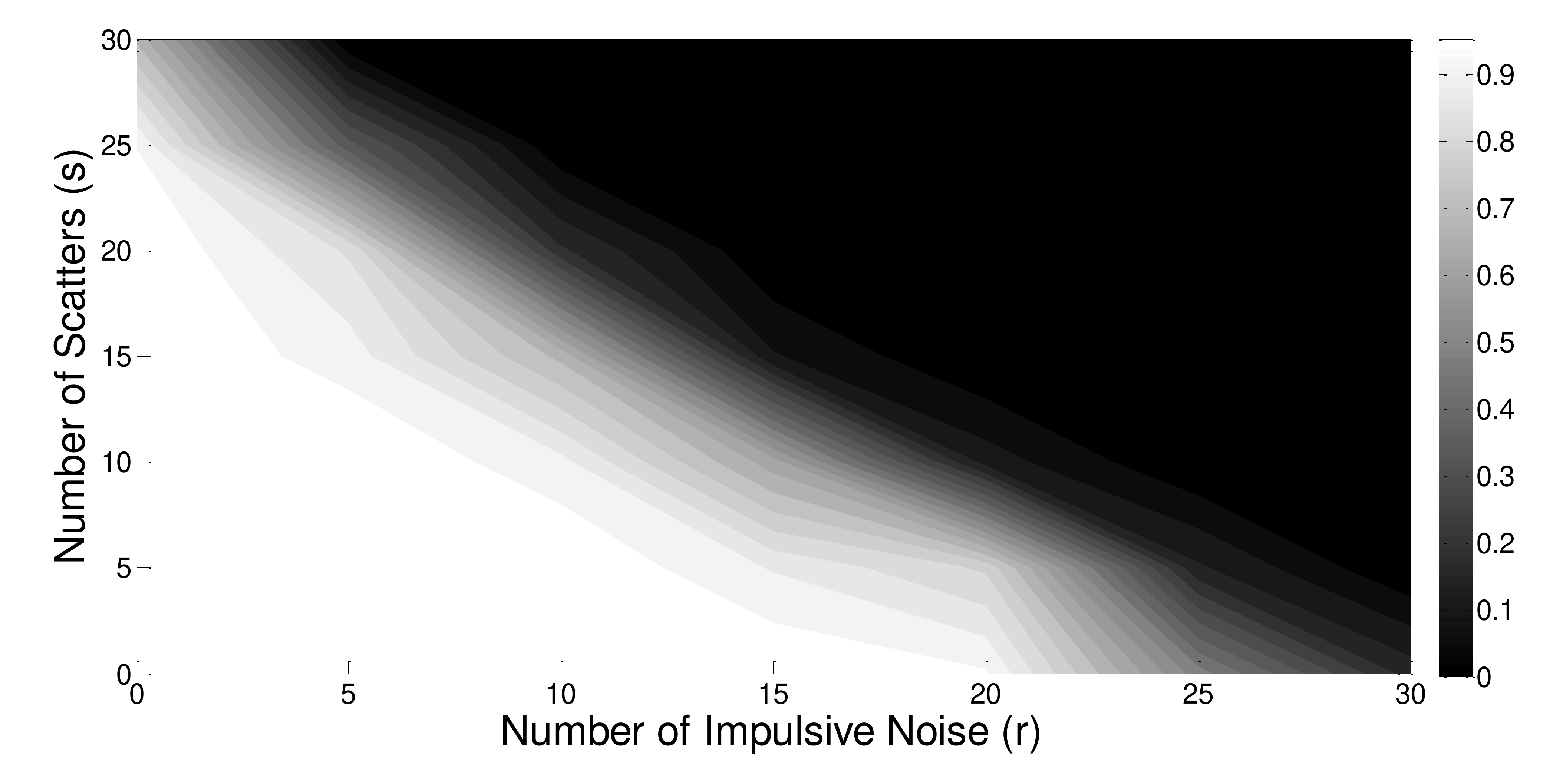}\label{fig.bound4}}
		\subfigure[]{\includegraphics[height=3cm, width=9.5cm]{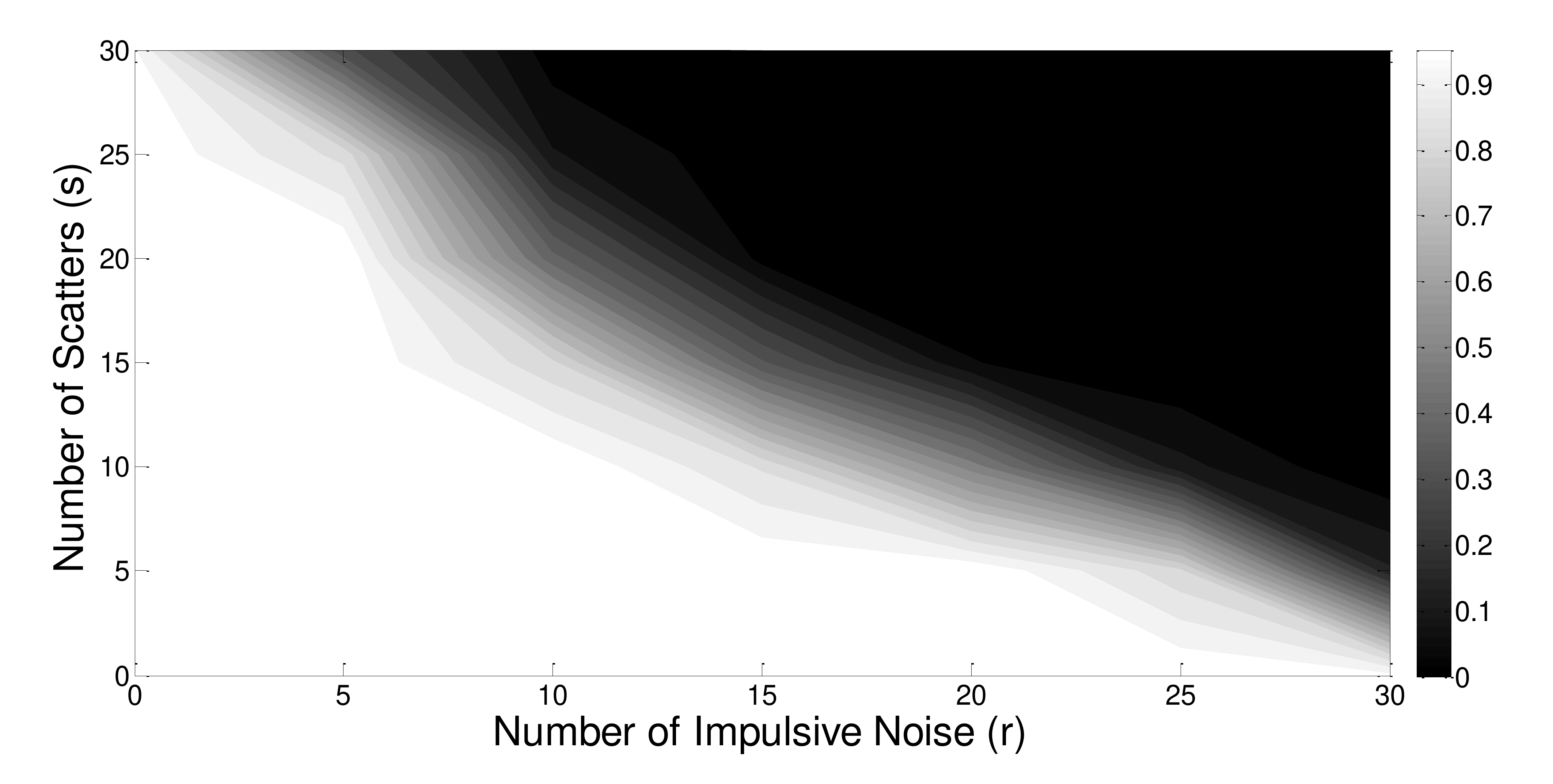}\label{fig.bound4}}
		}
	\vspace{-0.3cm}
	\caption{Phase transition graphs show the empirical rate of success of (\ref{sdp}) over $100$ trials. From the left to right $P=64$ and $P=128$, respectively. The white and black corresponding to perfect recovery and complete failure, respectively.\vspace{-0.8cm} }\label{fig.phase}.
\end{figure*}
The set of frequencies can be estimated by finding values of $f$ for which $|Q(f)|=1$ and the location of impulsive noise can be derived by $|\hat{\bm{q}}|=\lambda$, respectively (see Fig. \ref{fig.support}). When we obtain the set of ${\hat f_1,\cdots, \hat f_{\hat{s}}}$ for which $|Q(\hat f_k)| = 1$, each $\hat f_k$ is the reconstructed version of $f_k$, and $\hat{s}$ is the number of estimated sources, the corresponding complex coefficients of each path, $\hat{\alpha}_{k}$, can be achieved by the linear equations and LS algorithm 
$\label{gain}
\bm{h}=\sum_{k=1}^{\hat{s}}\alpha_{k}\bm{a}(\hat{f}_{k},0)
$.
 Eventually, the time dispersive channels can be calculated using $\hat{\tau}_{k}=\tfrac{PT_{s}}{\hat{f}_{k}}$.  
\section{Experiment} \label{section.simulation}
In this section, the performance of the proposed estimator (\ref{sdp}) was investigated by several MATLAB simulations. In the experiments, we randomly generated the time dispersive channels and impulsive noise on $[T_{s}, (P-1)T_{s})$ and $\mathcal{J}$, respectively. The gains of each paths were i.i.d. zero mean circularly symmetric complex Gaussian amplitude with unit variance.  Due to the fact that the low-pass polynomial (\ref{dualpoly}) was used to detect the inherent frequencies, the sparsity was not single-handedly sufficient and a minimum separation condition was needed, i.e., $|\tau_{k}-\tau_{j}| \geq 1.5T_{s}$ or $|f_{k}-f_{j}| \geq 1.5/P$\ for $i \neq j$ \cite{tang2013compressed}. At the receiver, we assumed accurate time-synchronization, $g(n)=1$ for all $n$, $f_{k}:= \tfrac{1}{P}\tfrac{\tau_{k}}{T_{s}},~\forall i=\{1,\cdots,s\}$, $\lambda=0.1$, $N=512$, $T_{s}=5 \mu s$, and the well-known pilot allocation scheme proposed in \cite{simeone2004pilot}.

 At the first experiment, we demonstrated that the locations where the magnitude of the dual polynomial (\ref{dualpoly}) achieves one and the magnitude of the dual solution achieves $\lambda$ are associated with the inherent frequencies in (\ref{measutmentmatrixform}) and the locations of impulsive noise, respectively. In this experiment,  $s=5$, $r=5$, signal-to-noise ratio \big($\text{SNR}=10\log_{10}\tfrac{1}{\sigma_{np}^{2}}$\big),  $\text{SNR}=10$dB, and $P=64$ were used. We carried out (\ref{sdp}) using CVX \cite{grant2008cvx} and obtained $\hat{\bm{q}}$. The dual polynomial then constructed by (\ref{dual}). Fig.\ref{fig.support} demonstrates frequencies and impulsive noise localization. In Fig. \ref{fig.fre}, the blue curve, the green and red lines are represented the dual polynomial, the support of the inherent and reconstructed frequencies, respectively. In Fig. \ref{fig.nois}, the dual solution and the location of impulsive noise are shown by the blue line and red stars, respectively. The gain of each frequencies can be calculated by (\ref{gain}). Then one can readily obtain the delays of each paths by $\hat{\tau}_{k}=\tfrac{PT_{s}}{\hat{f}_{k}}$.
 
\begin{figure}[t]
	\hspace{-1cm}
	\includegraphics[width= 10cm ,height=3.2cm]{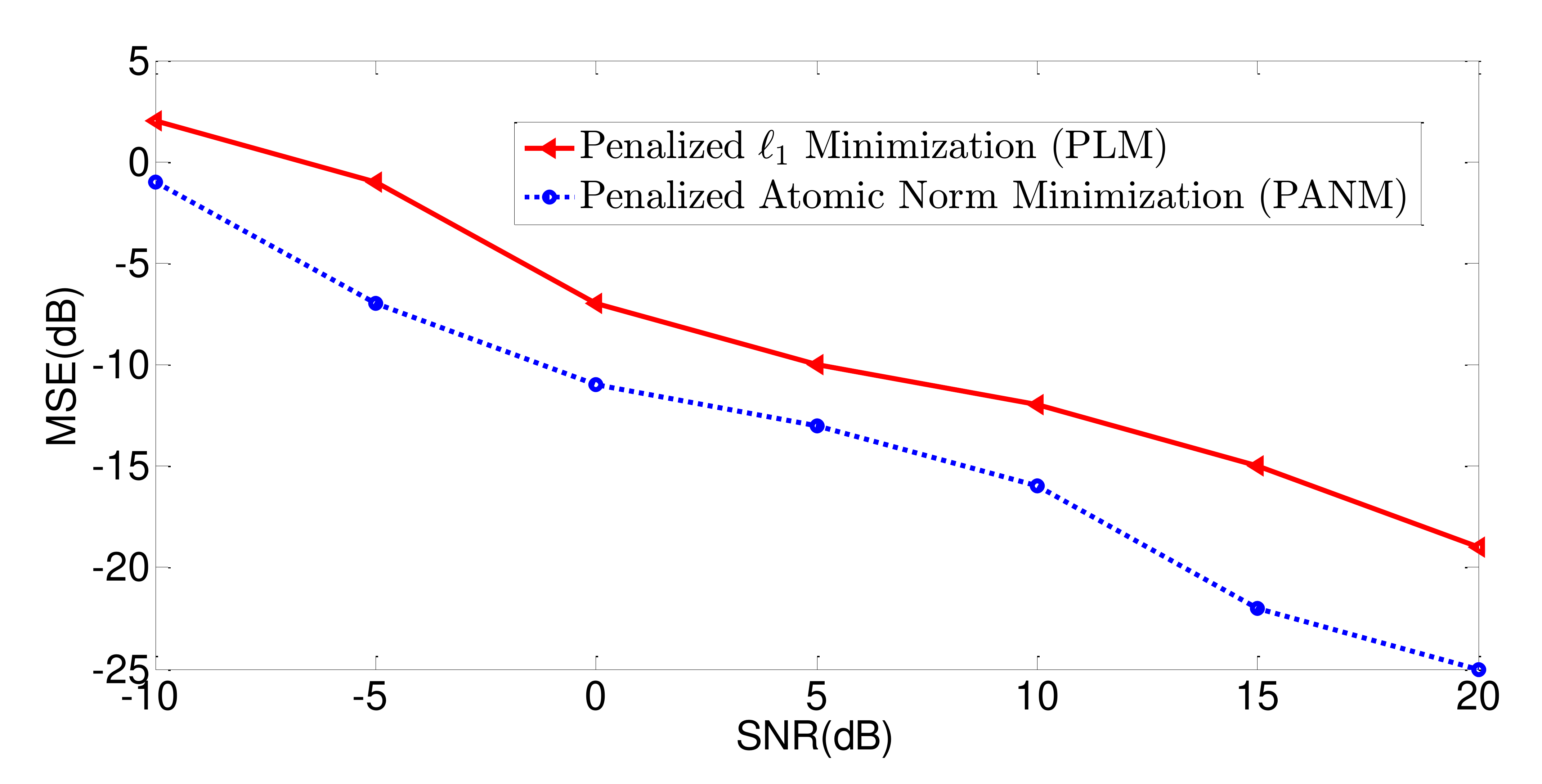}\vspace{-0.3cm}
	\caption{ The graph compares the MSE \big($\|\bm{h}-\hat{\bm{h}}\|_2$\big), where $\hat{\bm{h}}$ in the reconstructed channel, of the proposed estimator (\ref{sdp}) with the penalized LM \cite{popilka2007signal} over $100$ trials.}\vspace{-0.5cm}\label{fig.boundnew}
\end{figure}

At the second experiment, we evaluated the empirical rate of success of the proposed estimator (\ref{sdp}) using phase transition graphs. For each pair of modes $(r,s)$, we ran $100$ experiments. The experiment was called successful if the mean squared error (MSE) $\|\bm{h}-\hat{\bm{h}}\|_2 \le 10^{-2}$, where $\hat{\bm{h}}$ is the reconstructed channel. Fig. \ref{fig.phase} shows the success rate for each pair $(r,s)$ for $P=64$, $P=128$, and $\text{SNR}=30$dB. 

Finally, in Fig. \ref{fig.boundnew}, we examined MSE vs SNR of the proposed estimator (\ref{sdp}) to compare its performance with penalized $\ell_{1}$ minimization (PLM) \cite{popilka2007signal}. To prepare this experiment, we picked $P=128$, $r=6$, and $s=6$. Note that the regular LM \cite{cheng2013channel} and ANM \cite{pejoski2015estimation} are not able to estimate the parameters of the channel in this case because their optimization problems cannot model impulsive noise.  

\vspace{-0.1cm}
\section{Conclusion and Research Direction}\label{section.concultion}
	In this paper, a novel, simple estimator based on the PANM for estimating pilot-aided OFDM channel in an additive Gaussian and impulsive noise environment is proposed. It has shown that the dual of the PANM can be converted to a tractable semidefinite programming using PTP theory. Simulation results revealed that the proposed estimator considerably outperforms
	the PLM in terms of MSE. Due to the fact that CVX \cite{grant2008cvx} uses the interior point method with Newton
	steps, the complexity of implementing  (\ref{sdp}) is $\mathcal{O}(P^{3})$. We suggest developing an algorithm based the alternating direction method of multipliers (ADMM) in solving (\ref{sdp}) with faster running time as discussed in \cite{zheng2017super} for the OFDM passive radar.
	\vspace{-0.2cm}
\bibliographystyle{ieeetr}
\bibliography{HBReference}
\end{document}